\documentclass[reprint,aps,prd,superscriptaddress,showkeys,showpacs]{revtex4-1}
\usepackage{epsfig,amsmath,natbib}

\usepackage{aas_macros}
\usepackage{amssymb}
\usepackage{amsmath}
\usepackage{dsfont}
\usepackage{hyperref}
\usepackage{color}
\usepackage{pbox}
\usepackage{enumerate}

\hypersetup{
	colorlinks=false,
	citecolor=green
}

\begin{document}

%Heading********************************************************************************************************
\title{Do dark matter halos explain lensing peaks?}

\author{Jos\'e Manuel Zorrilla Matilla}
\email{jz2596@columbia.edu}
\affiliation{Department of Astronomy, Columbia University, New York, NY 10027, USA}
\author{Zolt\'an Haiman}
\affiliation{Department of Astronomy, Columbia University, New York, NY 10027, USA}
\author{Daniel Hsu}
\affiliation{Department of Computer Science, Columbia University, New York, NY 10027, USA}
\author{Arushi Gupta}
\affiliation{Department of Computer Science, Columbia University, New York, NY 10027, USA}
\author{Andrea Petri}
\affiliation{Department of Physics, Columbia University, New York, NY 10027, USA}

\date{\today}

\begin{abstract}
We have investigated a recently proposed halo-based model, \textsc{Camelus}, for predicting weak-lensing peak counts, and compared its results over a collection of 162 cosmologies with those from N-body simulations. While counts from both models agree for peaks with $\mathcal{S/N}>1$ (where $\mathcal{S/N}$ is the ratio of the peak height to the r.m.s. shape noise), we find $\approx 50\%$ fewer counts for peaks near $\mathcal{S/N}=0$ and significantly higher counts in the negative $\mathcal{S/N}$ tail. Adding shape noise reduces the differences to within $20\%$ for all cosmologies. We also found larger covariances that are more sensitive to cosmological parameters. As a result, credibility regions in the $\{\Omega_m, \sigma_8\}$ are $\approx 30\%$ larger. Even though the credible contours are commensurate, each model draws its predictive power from different types of peaks. Low peaks, especially those with $2<\mathcal{S/N}<3$, convey important cosmological information in N-body data, as shown in \cite{DietrichHartlap, Kratochvil2010}, but \textsc{Camelus} constrains cosmology almost exclusively from high significance peaks $(\mathcal{S/N}>3)$. Our results confirm the importance of using a cosmology-dependent covariance with at least a 14\% improvement in parameter constraints. We identified the covariance estimation as the main driver behind differences in inference, and suggest possible ways to make \textsc{Camelus} even more useful as a highly accurate peak count emulator.
\end{abstract}

\maketitle

% Introduction*****************************************************************************************************
\section{Introduction}
Weak gravitational lensing (WL) of background sources by large-scale structure (LSS) is a promising technique to study dark matter (DM) and dark energy (DE) \cite{DETF} as a consequence of its sensitivity to both structure growth and the expansion history of the universe.  Ongoing and future surveys such as the Dark Energy Survey (DES\footnote{\url{http://www.darkenergysurvey.org}}), the Euclid Mission\footnote{\url{http://sci.esa.int/euclid/}}, the Wide Field Infrared Survey Telescope (WFIRST\footnote{\url{http://wfirst.gsfc.nasa.gov}}) and the Large Synoptic Survey Telescope (LSST\footnote{\url{http://www.lsst.org}}) will deliver WL datasets with unprecedented precision, sky coverage and depth. For a comprehensive treatment of weak lensing in a cosmological context, we refer the reader to the following reviews \cite{BartelmannSchneider, Refregier03, Kilbinger15}.

On small scales, WL probes the matter density field in the non-linear regime, independent of the matter's nature or dynamic state. Thus, in order to optimally extract cosmological information from the upcoming WL surveys, we need observables that go beyond quadratic statistics such as the two-point correlation function or its Fourier transform, the power spectrum. Various strategies have been proposed to capture non-Gaussian information, from the use of higher-order moments and correlation functions such as the bispectrum (\cite{Bernardeau1997, Hui99, TakadaJain2003, Schneider2005}), to the adoption of topological features from WL maps such as Minkowski functionals \cite{Kratochvil2012, Petri2015} or peak counts \cite{Jain2000}.

Lensing peaks, defined as local maxima of the convergence or shear field, are particularly simple to extract from mass-aperture maps, and have been shown to constrain cosmology both theoretically \cite{DietrichHartlap, Kratochvil2010, Marian2012} and, recently, observationally \cite{Liu2015, LiuPan2015, DES2016}. Peaks are usually classified based on their absolute height or significance level, defined as their signal-to-noise ratio $(\mathcal{S/N})$, the noise being caused by our imperfect knowledge of the intrinsic shapes of the background galaxies.

Peak counts are also special because their physical origin and sensitivity to cosmology can, in principle, be understood and related to specific structures of the cosmic web. While our understanding is not yet complete, it is clear that halos are important contributors to peak counts. Shear peaks were initially considered for cluster selection, and the connection of high--significance peaks $(\mathcal{S/N}>4-5)$ to single massive halos has been established in the literature \cite{White2002, Hamana2004, Hennawi2005}. Lower--significance peaks are typically associated with constellations of lower-mass halos \cite{Yang2011,Liu2016} and contribute significantly to the cosmological information in convergence maps \cite{Kratochvil2010,Yang2011}. 

Predicting analytically the abundance of peaks is difficult, as it depends on projections of non-linear structures. N-body simulations can predict peak counts at a high computational cost that will only increase with the high volumes required by upcoming WL surveys. The need to predict not only the peak number density but also its covariance would further raise the total cost. The halo-peak connection has inspired some models that would circumvent the need for full N-body simulations by using either analytical models based on Gaussian random fields \cite{Fan2010,Maturi2010,Reischke2016} or stochastic fast simulations based on the halo model \cite{Kainulainen2009, LKI}. This could prove extremely useful by reducing the computational requirements for N-body simulations by 2-3 orders of magnitude.

The main goal of this work is to assess the validity of halo-based models for cosmological parameter inference. In particular, we compare results from full N-body simulations with those of a recent publicly available algorithm, \textsc{Camelus} \cite{LKI}. In previous work \cite{LKI}, this model was found to predict accurately peak counts from N-body simulations for a specific cosmology. Here, we expand the comparison of peak counts to a wide range of different cosmologies, and also examine their predicted covariance matrices, showing how differences affect the resulting parameter credibility regions. We also review the importance of the cosmology-dependence of the covariance matrix in the context of precision parameter inference \cite{LKII}.

The rest of the paper is organized as follows. In Sec.~\ref{methods} we describe the methods used to predict peak counts using N-body simulations and \textsc{Camelus}, and infer constraints for cosmological parameters. In Sec.~\ref{results} we show how both models compare in terms of peak counts, covariance matrices, and credible contours. We then discuss our main findings (Sec.~\ref{discussion}), identifying potential origins for the differences between the two models and how \textsc{Camelus} could be modified to match N-body predictions more accurately. Our main conclusions are summarized in Sec.~\ref{conclusions}.

% Predicting peak counts***************************************************************************************
\section{Predicting peak counts}\label{methods}
We generated convergence maps for a suite of 162 flat $\Lambda$CDM cosmologies covering the $\{\Omega_m, \sigma_8\}$ plane using both N-body simulations and \textsc{Camelus}. Table \ref{fiducial} presents the cosmological parameters for our fiducial cosmology, which are consistent with the 9-year Wilkinson Microwave Anisotropy Probe (WMAP) results \cite{WMAP} for ease of comparison with past simulation efforts.

\begin{table}
\caption{\label{fiducial}Cosmological parameters for the fiducial model. All other cosmologies share these parameters except $\Omega_m$ and $\sigma_8$.}
\begin{ruledtabular}
\begin{tabular}{lcc}
Parameter 							& Symbol 				& Value  \\
\hline
Matter density							& $\Omega_m$			& 0.260 \\
Dark energy density						& $\Omega_{\rm DE}$ 	& $1.0 - \Omega_m$ \\
Amplitude of fluctuations at 8 $h^{-1}$Mpc	& $\sigma_8$			& 0.800 \\
\hline
Hubble constant 						& $h$				& 0.72\\
Dark energy eq. of state					& $w$ 				& -1.0\\
Scalar spectral index 					& $n_s$ 				& 0.96\\
Effective number of relativistic d.o.f. 			& $n_{\rm eff}$			& 3.04\\
\end{tabular}
\end{ruledtabular}
\end{table}

We sampled the parameter space with a modified latin hypercube algorithm implemented in the publicly available lensing package \textsc{Lenstools} \cite{Petri2016}, and based on a coordinate transformation that converts a randomly sampled rectangle into an ellipse:
\begin{eqnarray}
(r,\phi) \rightarrow (x=ar^n\cos\phi, y=br^n\sin\phi) 
\end{eqnarray}
with $(r,\phi) \in [0,1] \times [0,2\pi]$. We adjusted the semi-axes $a$ and $b$ so that the region explored covered all areas with a significant likelihood according to past WL peak counts studies \cite{DietrichHartlap}. We centered the ellipse on our fiducial cosmology $(\Omega_m=0.260, \sigma_8=0.800)$, and rotated it so that its semi-major axis became parallel to the direction of maximum degeneracy between the two parameters. The exponent $n$ controls the sampling concentration, with $n>1/2$ yielding samples whose density grows towards the center of the ellipse. We used $n=3/2$.

Based on the likelihood estimated from a first batch of 100 cosmologies, we added manually 62 cosmologies in sparsely sampled regions, such as the contours' tails. Doing so reduced the sampling error in the likelihoods, as discussed in Sec.~\ref{discussion}.

\subsection{N-body simulations}\label{nbody}%***********************************************************
Our simulation pipeline is described in detail in \cite{Petri2016}. For each cosmology, we evolved a single $(240\, h^{-1} {\rm Mpc})^3$ volume with \textsc{Gadget2} \cite{Springel2005}, large enough to cover the intended $3.5\times3.5\,{\rm deg^2}$ field of view to a distance beyond the lensed sources' redshifts. Every simulated box contains $512^3$ DM particles,  which yields a mass resolution of $M_p\approx10^{10} {\rm M_{\odot}}$. All lensed source galaxies were placed at a redshift of $z_{\rm s}=1$, and $80\, h^{-1}{\rm Mpc}$ thick lens planes were stacked between the galaxies and the observer. Each lens plane is the result of slicing a snapshot along a coordinate axis, and applying to it a random shift and rotation, allowing us to generate 500 independent realizations from a single N-body run. Lens planes were converted to potential planes and a multi-plane ray-tracing algorithm was used to generate $1,024\times1,024$ pixels convergence maps with a pixel size of $\approx 0.2 \, {\rm arcmin}$. We used a higher resolution for the potential planes, $4,096 \times 4,096 $ pixels, to avoid a loss of power on small scales \cite{Yang2011}. We deployed and managed the simulations and their output using \textsc{Lenstools} \cite{Petri2016}.

Since the unperturbed galaxy shape is unknown, we accounted for an intrinsic ellipticity noise following \cite{VanWaerbeke2000} and added a 2-D Gaussian random noise with zero mean and standard deviation
\begin{eqnarray}
\sigma_{pix} = \sqrt{\frac{\sigma_{\epsilon}^2}{2 n_g A_{pix}}}
\end{eqnarray}
with intrinsic ellipticity $\sigma_{\epsilon}=0.4$ as in \cite{LKI}, a galaxy density of $n_g=25 \, {\rm arcmin}^{-1}$ and pixel area defined by the field-of-view and map resolution. We smoothed the noiseless and noise-only maps applying a Gaussian filter with a characteristic width of $\theta_G=1$ arcmin --see Eq.~\ref{window}-- before combining them, and extracted their local maxima, recording them in the form of peak catalogues.

\begin{eqnarray}
W(\theta) = \frac{1}{\pi \theta_G^2} \exp(-\frac{\theta^2}{\theta_G^2})
\label{window}
\end{eqnarray}

\subsection{Camelus}%**************************************************************************************
\textsc{Camelus} is a halo-based model that generates fast stochastic simulations of convergence maps. Instead of evolving the matter density field from high redshift dynamically, it assumes that halos are the primary contributors to the lensing signal and discretizes the space between the lensed galaxies and the observer in redshift bins, populating them with halos whose masses are sampled from an analytical function \cite{Jenkins2001}. Each halo follows a Navarro-Frenk-White (NFW) density profile \cite{Navarro1997} and is placed randomly within its redshift bin. We refer the reader to \cite{LKI} for an in-depth description of the model.

We ran \textsc{Camelus} for each of the same set of 162 cosmologies as with the N-body simulations, generating 500 independent  realizations in each case. The resulting smoothed, noiseless convergence maps were combined with shape noise that is statistically the same as the one used with the N-body maps, and their peaks extracted with the same routines. The values we used for the relevant tunable parameters in \textsc{Camelus } are given in Table \ref{CamelusPar}.

\begin{table}
\caption{\label{CamelusPar} The main tunable parameters of \textsc{Camelus} and their values used in this study.  Dark matter halos are assumed to have a Navarro-Frenk-White (NFW) density profile, defined by its inner slope $(\alpha)$, and its concentration parameter, the ratio between the virial and scale radii, determined itself by $c_0$ and $\beta$: $c_{\rm NFW}\equiv \frac{c_0}{1+z}\left(\frac{M}{M_{\star}} \right)^{\beta}$, where $z$ is the halo's redshift, $M$ its mass and $M_{\star}$ its pivot mass (see \cite{LKI} for a detailed description of the halo density profile characterization).}
\begin{ruledtabular}
\begin{tabular}{lcc}
Parameter 			& Symbol 				& Value  \\
\hline
Field of view 			& $fov$				& $210.0\times210.0\, {\rm arcmin}^{2}$\\
Pixel size 				& - 					& $0.205 \, {\rm arcmin}$\\
Smoothing scale 		& $\theta_G$ 			& $1.0\,{\rm arcmin}$\\
\hline
Minimum halo mass 		& $M_{min}$ 			& $10^{11}\,h^{-1} {\rm M_{\odot}}$ \\
Maximum halo mass 	& $M_{max}$ 			& $10^{17}\,h^{-1}{\rm M_{\odot}}$ \\
Maximum halo redshift 	& $z_{max}$ 			& 1.0 \\
No. of redshift bins 	& $n_z$ 				& 10 \\
Halo profile inner slope 	& $\alpha$  			& 1.0\\
Halo concentration (norm.)	& $c_0$ 				& 11.0 \\
Halo concentration (slope)	& $\beta$ 				& 0.13 \\
\hline
Galaxies redshift 		& $z_{gal}$ 			& 1.0 \\
Galaxy density 			& $n_g$ 				& $25.0 \, {\rm arcmin}^{-2}$\\
Ellipticity noise 			& $\sigma_{\epsilon}$ 	& 0.4 \\
\end{tabular}
\end{ruledtabular}
\end{table}

% Inference******************************************************************************************************
\subsection{Parameter inference}\label{inference}
Bayes' theorem relates the probability distribution for a set of cosmological parameters, given an observation, to the likelihood of the observed data given values for those parameters
\begin{eqnarray}
p(\theta | \mathbf{x}^{obs},M)=\frac{p(\mathbf{x}^{obs} | \theta,M)p(\theta,M)}{p(\mathbf{x}^{obs},M)}
\end{eqnarray}
where $p$ is the probability, $\theta$ represents the set of parameters that determine the model $M$ and $\mathbf{x}^{obs}$ is a data vector that depends on observations. Throughout this study we assume $\Lambda$CDM is a correct description of the universe, hence the evidence (denominator) acts just as a normalizing factor and we can drop the implicit dependence on the model. We use a non-zero prior within the parameter region that we explore, and zero outside:
\begin{eqnarray}
p(\theta | \mathbf{x}^{obs}) \propto p(\mathbf{x}^{obs} | \theta)\equiv \mathcal{L}(\theta) 
\end{eqnarray}
Our observable is the peak function defined as the peak counts binned by their height or significance level ($\mathcal{S/N}$, height in units of the  r.m.s. ellipticity noise). 

If we assume that our observable follows a multivariate Gaussian distribution, its log-likelihood, up to an additive constant, has the form:
\begin{eqnarray}
&L_{vg} = \ln \left[(2d)^d \det C(\theta) \right] + \Delta \mathbf{x}^T(\theta) \widehat{C^{-1}}(\theta) \Delta \mathbf{x}(\theta)
\label{Lvg}
\end{eqnarray}
where $\Delta \mathbf{x}$ is the difference between the mean peak function in each cosmology from its value in the fiducial $(\Omega_m=0.260, \sigma_8=0.800)$ cosmology, and $\widehat{C^{-1}}$ is the precision matrix (the inverse of the covariance matrix), estimated from the data. We follow the same notation as \cite{LKII} , and call it $L_{vg}$, $L$ indicating it is a "log-likelihood", $v$ that it includes a "varying" (i.e. cosmology-dependent) covariance matrix, and $g$ that the assumed model is "Gaussian".

Means and covariance matrices are computed from the $N=500$ realizations available in each cosmology:
\begin{eqnarray}
&\Delta \mathbf{x}(\theta) = \bar{\mathbf{x}}(\theta) - \bar{\mathbf{x}}(\theta_{fid})
\end{eqnarray}
\begin{eqnarray}
&C(\theta) = \frac{1}{N-1} \sum\limits_{i=1}^{N}(\mathbf{x}_i(\theta)-\bar{\mathbf{x}}(\theta))(\mathbf{x}_i(\theta)-\bar{\mathbf{x}}(\theta))^T
\label{eqcov}
\end{eqnarray}
In many cases, evaluating the covariance matrix at each point of the parameter space becomes computationally too expensive, and a constant covariance is used instead. As in \cite{LKII}, we assess the effect of this simplification by evaluating two approximations to the full Gaussian likelihood. The first is to use a "semi-varying" covariance matrix; i.e., we let the covariance matrix change with cosmology within the $\chi^2$ term but not the determinant term in Eq.~\ref{Lvg}. Following the notation in \cite{LKII} we call it $L_{svg}$. The second is to compute the likelihood with a "constant" covariance matrix, evaluated at the fiducial model, in all terms. We call this $L_{cg}$:
\begin{eqnarray}
&L_{svg} = \Delta \mathbf{x}^T(\theta) \widehat{C^{-1}}(\theta) \Delta \mathbf{x}(\theta) 
\label{Lsvg}\\
&L_{cg} = \Delta \mathbf{x}^T(\theta) \widehat{C^{-1}}(\theta_{fid}) \Delta \mathbf{x}(\theta) 
\label{Lcg}
\end{eqnarray}
Note that the precision matrices in Eqs.~\ref{Lvg}, \ref{Lsvg} and \ref{Lcg} have a "hat" on top, while the covariance matrix in Eq.~\ref{eqcov} does not. That is because the inverse of a covariance matrix estimated from data is not an unbiased estimator for the precision matrix. There are two ways to correct for the bias. The most common \cite{Hartlap2007} is to rescale the inverse of the estimated covariance matrix: 
\begin{eqnarray}
\widehat{C^{-1}} = \frac{N-d-2}{N-1}C^{-1}
\end{eqnarray}
where N is the number of realizations per cosmology (500 in our case) and $d$ the dimension of the observable (number of bins in the peak function).

An alternative approach is to use a non-Gaussian likelihood, as described in  \cite{Sellentin}. In this case we can also use a constant or varying covariance matrix in each of the log-likelihood terms and, following the same notation, drop the $g$ subscript since the model is not a Gaussian anymore. The functional form for these models is as follows:
\begin{eqnarray}
&L_{v}=\ln \left[\frac{\det C(\theta)}{c_p^2}\right] + N\left[1+\frac{\Delta\mathbf{x}^T (\theta)C^{-1}(\theta)\Delta\mathbf{x}(\theta)}{N-1}  \right]\\
&L_{sv}=N\left[1+\frac{\Delta\mathbf{x}^T (\theta)C^{-1}(\theta)\Delta\mathbf{x}(\theta)}{N-1}  \right]\\
&L_{c}=N\left[1+\frac{\Delta\mathbf{x}^T (\theta)C^{-1}(\theta_{fid})\Delta\mathbf{x}(\theta)}{N-1}  \right]\\
\end{eqnarray}
with a normalizing factor
\begin{eqnarray}
\bar{c}_p = \frac{\Gamma \left(\frac{N}{2} \right)}{\left[ \pi(N-1)\right]^{d/2}\Gamma \left( \frac{N-d}{2} \right)}
\end{eqnarray}
where $\Gamma$ is the usual Gamma function and $N>d$. 

In the limit $N \gg d$ both methods are equivalent. We used peak functions with a relatively small number of bins (see below) compared with the number of realizations per model and there were no discernible differences between the credible contours generated using the two approaches.

For inference, we decided to use few bins in the peak function so that covariance bias is not an issue. We set an edge at $\mathcal{S/N}=3.0$, the threshold below which peak counts are dominated by noise. This allowed us to separate clearly analyses done with only high-significance peaks (as in \cite{LKII}) from analyses also including low-significance and even negative peaks. The upper and lower $\mathcal{S/N}$ edges were chosen to avoid the rejection of models due to the presence of empty bins with their corresponding singular covariance matrices. We also ensured that there are at least 10 peaks from the fiducial cosmology in the bin with the lowest number and defined the 10-bin peak function described in Table \ref{bins}, $\mathbf{x}^{obs}\equiv\mathbf{n_{\rm pk}^{(10)}}$, as the observable for this study. We did not optimize the bins' edges to maximize the predictive power of the models.

Table \ref{bins} also displays $\mathbf{n_{\rm pk}^{(100)}}$, a peak function with 100 equally spaced bins that was used to highlight differences in peak counts from the two models.

\begin{table}
\caption{\label{bins}Description of the thresholds used in this study to bin the convergence peak counts by their signal-to-noise ($\mathcal{S/N}$) ratio, as well as the mean peak counts from data obtained from both the N-body and the \textsc{Camelus} models in the fiducial cosmology, in the bins used for inference.}
\begin{ruledtabular}
\begin{tabular}{lcc}
Observable 		& $\mathcal{S/N}$ bins \\				
\hline
$\mathbf{n_{\rm pk}^{(100)}}$ 	& 100 equally-sized bins in [-2.0, ..., 6.0]\\
$\mathbf{n_{\rm pk}^{(10)}}$ 	& $[-\infty, -1.0, 0.0, 1.0, 2.0, 3.0,$ \\
				& $3.5, 4.0, 4.5, 5.0, +\infty]$\\
\hline
Model			& $\mathbf{\bar{n}_{\rm pk}^{(10)}}$\\
\hline
N-body			& [23.8, 292.5, 1125.7, 1457.3, 735.4, \\
				&130.5, 59.8, 27.8, 13.3, 17.7]\\
\textsc{Camelus}	& [15.3, 255.6, 1145.8, 1535.1, 721.7, \\
				& 113.3, 48.1, 21.2, 10.4, 15.7\\
\end{tabular}
\end{ruledtabular}
\end{table}

We are forced to interpolate for all the $(\Omega_m, \sigma_8)$ combinations not found in our collection of simulated cosmologies in order to compute smooth credible contours. Our interpolation grid covers the region $\Omega_m \in [0.160,0.600]$ and $\sigma_8 \in [0.150,1.250]$ with a resolution of 0.001 on each axis. Within that region we know that our sample reproduces $2\sigma$ (95.4\%) contours from \textsc{Camelus} within 20\% --see Sec.~\ref{discussion}--, and we verified that a finer grid did not change the results. 

Interpolating peak counts is straightforward, and can be done when using a constant covariance to calculate the likelihood, but becomes problematic when an estimation for the covariance matrix is also needed. We interpolated the log-likelihood instead, and used a linear model because its results are easy to interpret, it does not require any tunable parameter like smoothing, and it does not introduce any spurious high-likelihood values from fitting high-order polynomials. We verified that our results do not change when using a different interpolator, such as radial basis functions; this agrees with the findings in previous studies such as \cite{DES2016}. 

% Results*******************************************************************************************************
\section{Results}\label{results}
Our main results are the comparison between the two models regarding peak counts, covariance matrices and credible contours, together with the impact of using a cosmology-dependent covariance for inference.

\begin{figure*}
\begin{center}
\includegraphics[width=1.0\textwidth]{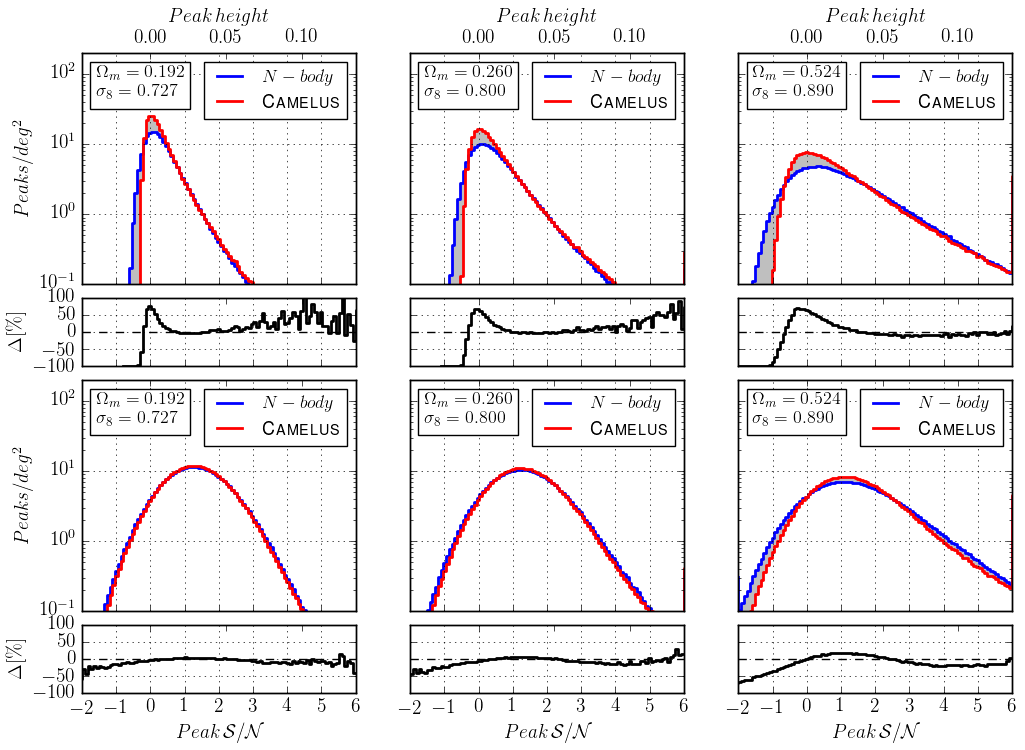}
\end{center}
\caption{
Comparison of mean peak counts as a function of their height between N-body simulations (blue) and \textsc{Camelus} (red). Counts are normalized to 1 deg$^2$ of sky and height is expressed in absolute value and as a signal-to-noise ratio $(\mathcal{S/N})$. The upper panels show the results from smoothed convergence maps without shape noise; the lower panels add shape noise. Three different cosmologies are displayed with increasing parameter $\Sigma_8$ from left to right (0.556, 0.734 and 1.244). In black, we show the fractional difference between the two models $(\Delta [\%] \equiv (N_{\textsc{Camelus}}-N_{N-body})/N_{Nbody})$, and the area between the histograms is shaded.
Adding noise reduces the discrepancies between the models but the effect depends on cosmology. While the discrepancies are almost erased for cosmologies with small $\Sigma_8$, for the rest N-body data yield lower counts near the maximum of the distribution and higher counts in the tails. The differences grow with $\Sigma_8$.
}
\label{histograms}
\end{figure*}

Fig.~\ref{histograms} shows mean peak counts as a function of their height, with and without galaxy shape noise, for three representative cosmologies that are characterized by the degeneracy parameter defined as in \cite{DES2016}, $\Sigma_8\equiv\sigma_8\left(\frac{\Omega_m}{0.3}\right)^{0.6}$. In each cosmology, we calculated the average of the peak function,  $\mathbf{\bar{n}_{\rm pk}^{(100)}}$, over 500 smoothed maps generated with the two models. We did this before and after adding noise as described in \ref{nbody}. Noiseless maps from N-body simulations exhibit up to 50\% fewer peaks around the maximum of the distribution at $\mathcal{S/N}\sim0$, with higher counts in the tails. Nevertheless, the two models agree well for peaks with $\mathcal{S/N}>1$, which constrain cosmology the most (see below). The addition of noise dilutes the differences for low-significance counts, especially for cosmologies with small $\Sigma_8$, and has the opposite effect for high-significance peaks, with N-body noisy maps yielding more counts for $\mathcal{S/N}>3$, especially for cosmologies with high $\Sigma_8$. 

%%%Figure

\begin{figure}
\begin{center}
\includegraphics[width=0.5\textwidth]{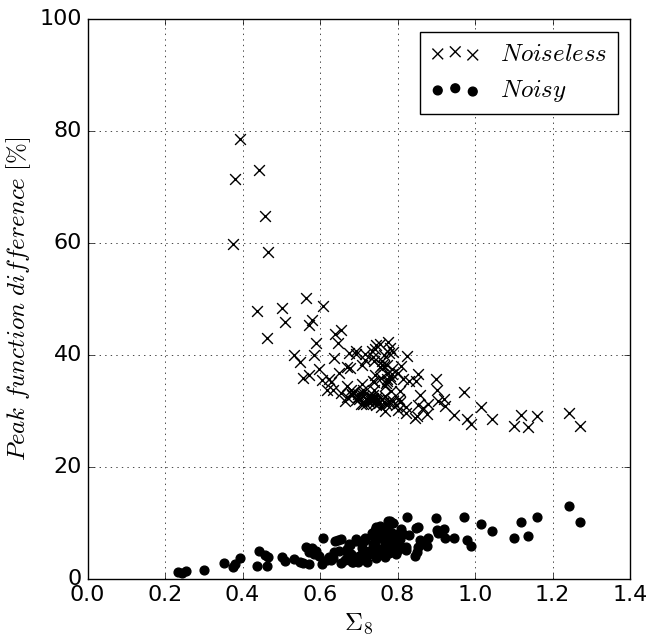}
\end{center}
\caption{Global comparison of peak counts. For each cosmology, the area between the N-body and \textsc{Camelus} histograms as a percentage of the area enclosed by the N-body histogram (Eq.~\ref{performanceMeasure}) is plotted against $\Sigma_8$. Differences from noiseless maps (crosses) are significantly reduced by adding noise (dots), so that the difference stays below 20\% in all cases. The reduction is more important for cosmologies with small $\Sigma_8$, for which noise dominates.}
\label{MeanPerformance}
\end{figure}

As a global measure of how different the peak histograms from the two models are, we integrated the area between them, divided the result by the surface under the N-body histograms --see Eq.~\ref{performanceMeasure}--, and plotted it as a function of $\Sigma_8$ in Fig.~\ref{MeanPerformance}. 
 
\begin{eqnarray}\label{performanceMeasure}
\Delta (\%) \equiv \frac{\sum\limits_{i=1}^{N_{bins}} |Npeaks_{\textsc{Camelus}}^i-Npeaks_{N-body}^i|}{\sum\limits_{i=1}^{N_{bins}} Npeaks_{N-body}^i}
\end{eqnarray}
 
Counts from noiseless convergence maps generated with the halo-based model are in better agreement with those from N-body simulations as $\Sigma_8$ increases, pointing to a higher non-halo contribution to peaks for small $\Sigma_8$ cosmologies. Adding noise reduces the global differences to less than 20\% in all cases. As expected, the reduction is stronger for cosmologies with small $\Sigma_8$ where peak counts are dominated by noise. Thus, the agreement between models worsens as $\Sigma_8$ increases. 

%%%Figure

\begin{figure}
\begin{center}
\includegraphics[width=0.5\textwidth]{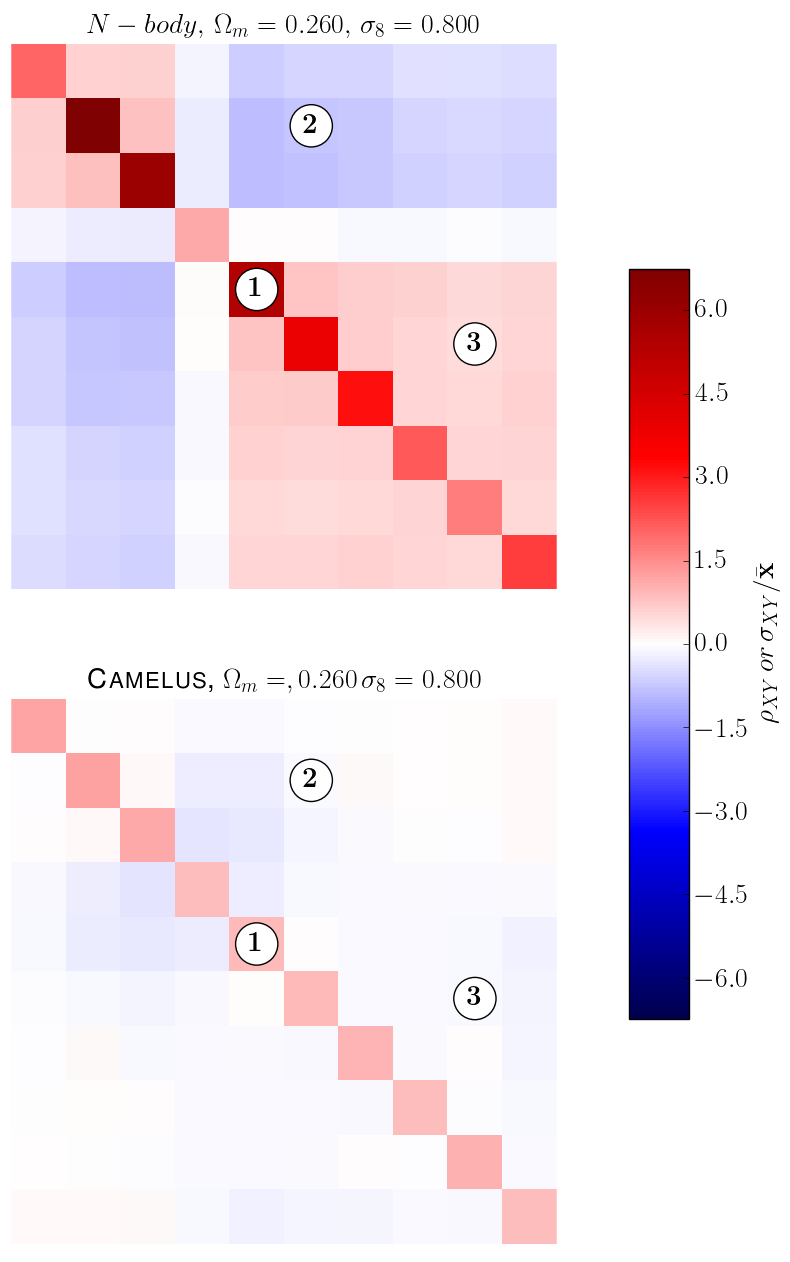}
\end{center}
\caption{Covariance comparison between N-body (upper panel) and \textsc{Camelus} (lower panel) for the fiducial cosmology. Each normalized covariance matrix has diagonal elements equal to the peak count variance divided by its mean, $\frac{\sigma_{ii}^2}{\bar{x}_{ii}}$, and off-diagonal elements equal to the correlation coefficients, $\rho_{ij}\equiv\frac{\sigma_{ij}}{\sigma_i\sigma_j}$. We find higher absolute values for all elements in the matrices, with positive and negative peaks positively correlated and positive peaks anti-correlated with negative ones. Peak counts from \textsc{Camelus} are mildly anti-correlated. Selected matrix elements whose value for all cosmologies is displayed in Fig. ~\ref{CovCosmo} are indicated with a number.}
\label{cov}
\end{figure}

Calculating the likelihood of a cosmological model needs an estimate of the covariance matrix, as seen in Sec.~\ref{inference}. We analyzed the covariances for $\mathbf{n_{\rm pk}^{(10)}}$, the data vector used to draw the credible contours. Fig.~\ref{cov} shows this comparison for the fiducial cosmology. Specifically, we display the correlation matrices after substituting their diagonal terms with the variances divided by the mean peak counts. These normalized matrices allow for a comparison of the variance and correlations for each bin, irrespective of its mean peak count. 

N-body data yield higher absolute values in all matrix elements. Positive and negative peaks have higher correlations among themselves, while being anti-correlated against one another. \textsc{Camelus} data, on the other hand, gives weakly anti-correlated peak counts with a smaller variance. The weak anti-correlation in the \textsc{Camelus} data can be attributed to the condition that the total mass in all halos is fixed: lens planes including an unusually large number of massive halos will have room for fewer low-mass halos, and vice-versa. Also, as we discuss in Sec.~\ref{discussion}, the covariance underestimation can be the consequence of halos being randomly placed in the field of view.

%%%Figure

\begin{figure*}
\begin{center}
\includegraphics[width=1.0\textwidth]{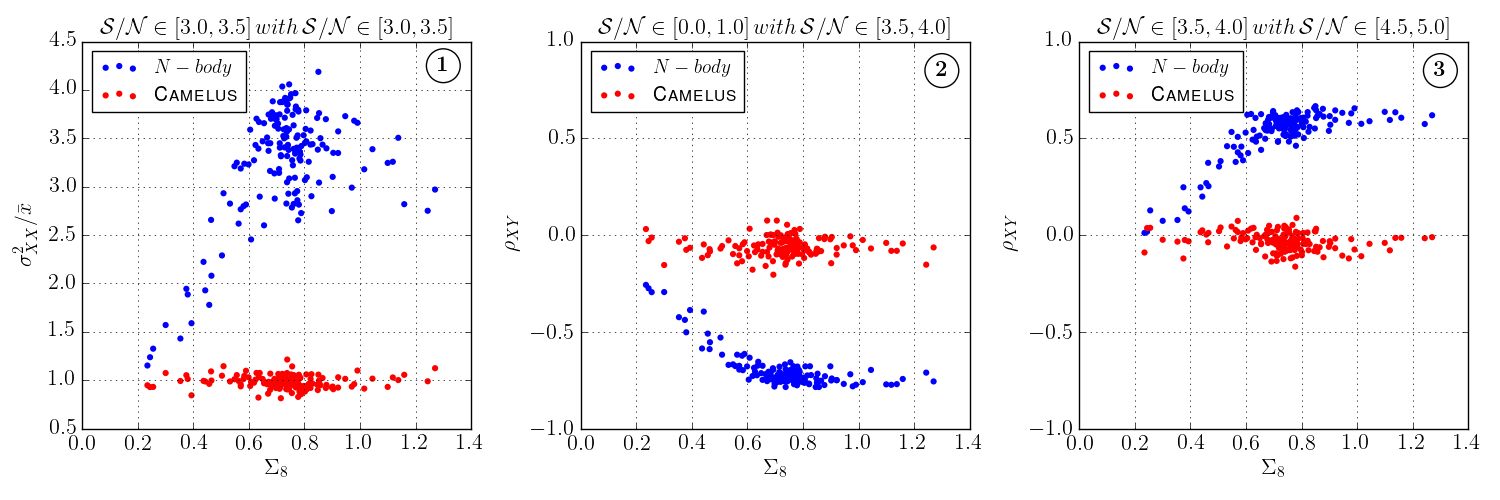}
\end{center}
\caption{The cosmology dependence of covariances. Each subplot shows the value of selected normalized covariance matrix elements for all 162 cosmologies. The selected elements are indicated in Fig. ~\ref{cov}, and correspond to a diagonal element (left panel) and off-diagonal elements showing anti-correlation in N-body data (center panel) and correlation (right panel). N-body data exhibit higher absolute values for all elements and stronger cosmology dependence.}
\label{CovCosmo}
\end{figure*}

To analyze the cosmology dependence of the covariance, we plotted the value of selected normalized matrix elements as a function of $\Sigma_8$ for all cosmologies in Fig.~\ref{CovCosmo}. For N-body data, all variances and correlations increase until $\Sigma_8\approx 0.6$ and then plateau. This dependence may affect the likelihood calculations. Matrices computed with \textsc{Camelus} show a very weak cosmology dependence and all their elements are smaller --in absolute value-- than those from N-body simulations, which would result in lower error estimations.  

%%%Figure

\begin{figure}
\begin{center}
\includegraphics[width=0.5\textwidth]{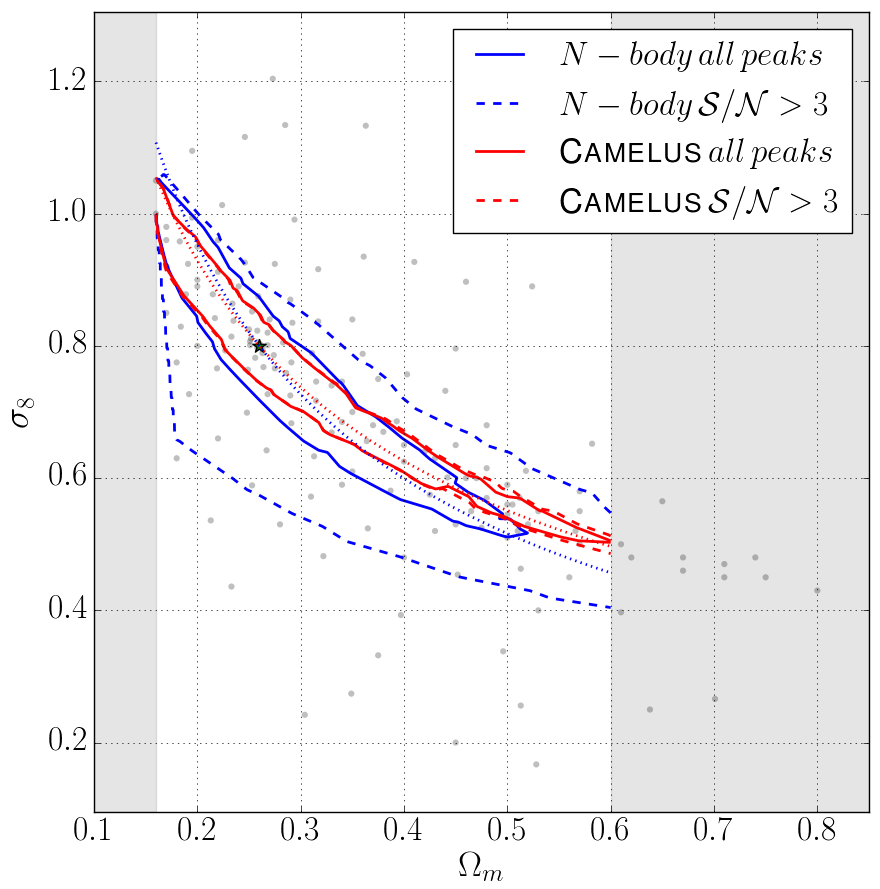}
\end{center}
\caption{Comparison of $2\sigma$ (95.4\%) credible contours from N-body (blue) and \textsc{Camelus} (red) data, using a Gaussian likelihood with constant covariance, $L_{cg}$. Solid lines show the contours computed using all the peak counts. We find looser constrains, with a thicker, $\approx 30\%$ larger credibility region. Dashed lines show the results including only high significance peaks $(\mathcal{S/N}>3)$. While constraints based on \textsc{Camelus} data do not change, the predictive power from N-body data is severely reduced, with a $\approx 200\%$ increase in the area of the credibility region. Dotted lines show the degeneracies $\Sigma_8 = \sigma_8 \left( \frac{\Omega_m}{0.3}\right)^\alpha$ that minimize scatter in $L_{cg}$. We find a steeper contour, $\alpha=0.67$ vs. $\alpha=0.58$ for \textsc{Camelus}.  Grey dots show the simulated cosmologies (a green star the fiducial cosmology), and grey areas the regions excluded from contour measurements.}
\label{contours1}
\end{figure}

After comparing peak counts and their covariances, we combined these to estimate the $L_{cg}$ likelihood for each model. We show the $2\sigma$ (95.4\%) credible contours in Fig.~\ref{contours1} by numerically integrating the interpolated likelihoods, and compared the results in Table ~\ref{infTable}. We find thicker contours, with a $30\%$ larger overall area, which can be attributed to the larger covariances (see below).

We also report any shifts in the credibility region's centroid position in Table ~\ref{infTable}. The centroid is defined as the point whose position is the arithmetic mean of that of all points within the region:
\begin{eqnarray}
\theta^{centroid} = \frac{\int_{CR} d\theta d\Theta \theta}{Area_{CR}} \approx \frac{\sum_{CR}\theta^i}{\sum_{CR}1}
\end{eqnarray}
where $\theta$ refers to the axis for which the centroid coordinate is computed and $\Theta$ to all other dimensions in parameter space. We did not use the maximum likelihood to estimate shifts because it corresponds to the fiducial cosmology by construction. We found a significant shift exclusively between N-body contours computed using all the peaks and those computed using only high-significance peaks. The contours from N-body simulations are more tilted in $\{\Omega_m,\sigma_8\}$. To quantify the difference in tilt, we fitted the exponent $(\alpha)$ of the degeneracy relation, $\Sigma_8 \equiv \sigma_8 \left( \frac{\Omega_m}{0.3}\right)^\alpha$ to minimize the scatter in $L_{cg}$. We restricted the data to $\Sigma_8 \in [0.6,0.9]$, since estimating the scatter for extreme values of $\Sigma_8$ where we have few data points is problematic. We find an exponent of $\alpha = 0.67$ vs. $\alpha=0.58$ for \textsc{Camelus}.

It is common to restrict analyses to the highly significant peaks, since their counts are not dominated by shape noise. We emphasize that the shape noise can be measured accurately from the data themselves, and so there is no reason a priori to discard the  'noisy' peaks with a lower $\mathcal{S/N}$. Nevertheless, we investigated the impact of this restriction. We find that it does not change the contours obtained with \textsc{Camelus}, but has a drastic impact on those from N-body simulations, as can be seen in Fig.~\ref{contours1}. Previous works (\cite{DietrichHartlap, Kratochvil2010}) found that low-significance peaks carry important cosmological information in WL maps from N-body simulations. Table \ref{infTable} shows that the contours double in size when only peaks with $\mathcal{S/N}>3$ are considered. While both models yield similar constraints, they derive their predictive power from different $\mathcal{S/N}$ peaks.

\begin{table}
\caption{\label{infTable}Comparison of $L_{cg}$ $2\sigma$ (95.4\%) credible contours. The figure-of-merit, $FoM$, is the inverse of the area of the credibility regions. Also displayed are the percentage changes in the area of the credibility region and its centroid (arithmetic mean) shift. We find looser constraints  $(\approx 30 \%)$ for N-body data, whose predictive power is greatly diminished when low significance peaks are excluded from the analysis.}
\begin{ruledtabular}
\begin{tabular}{lcccc}
				& $FoM$ 	&$\Delta\,Area$ 	& $\Delta \Omega_m$ 	& $\Delta \sigma_8$ \\
\hline
$N-body\,all\,peaks$ 				& 26 		& - 			& - 			& -\\
$N-body\,\mathcal{S/N}>3.0$ 			& 9 		& +198\% 		& +0.05		& -0.09\\
\hline
\textsc{Camelus} all peaks  			& 36  	& -28\% 		& +0.02 		& -0.00\\
\textsc{Camelus} $\mathcal{S/N}>3.0$ 	& 33 	  	& -21\%		& +0.03		& -0.02\\
\end{tabular}
\end{ruledtabular}
\end{table}

%%%Figure

\begin{figure*}
\begin{center}
\includegraphics[width=1.0\textwidth]{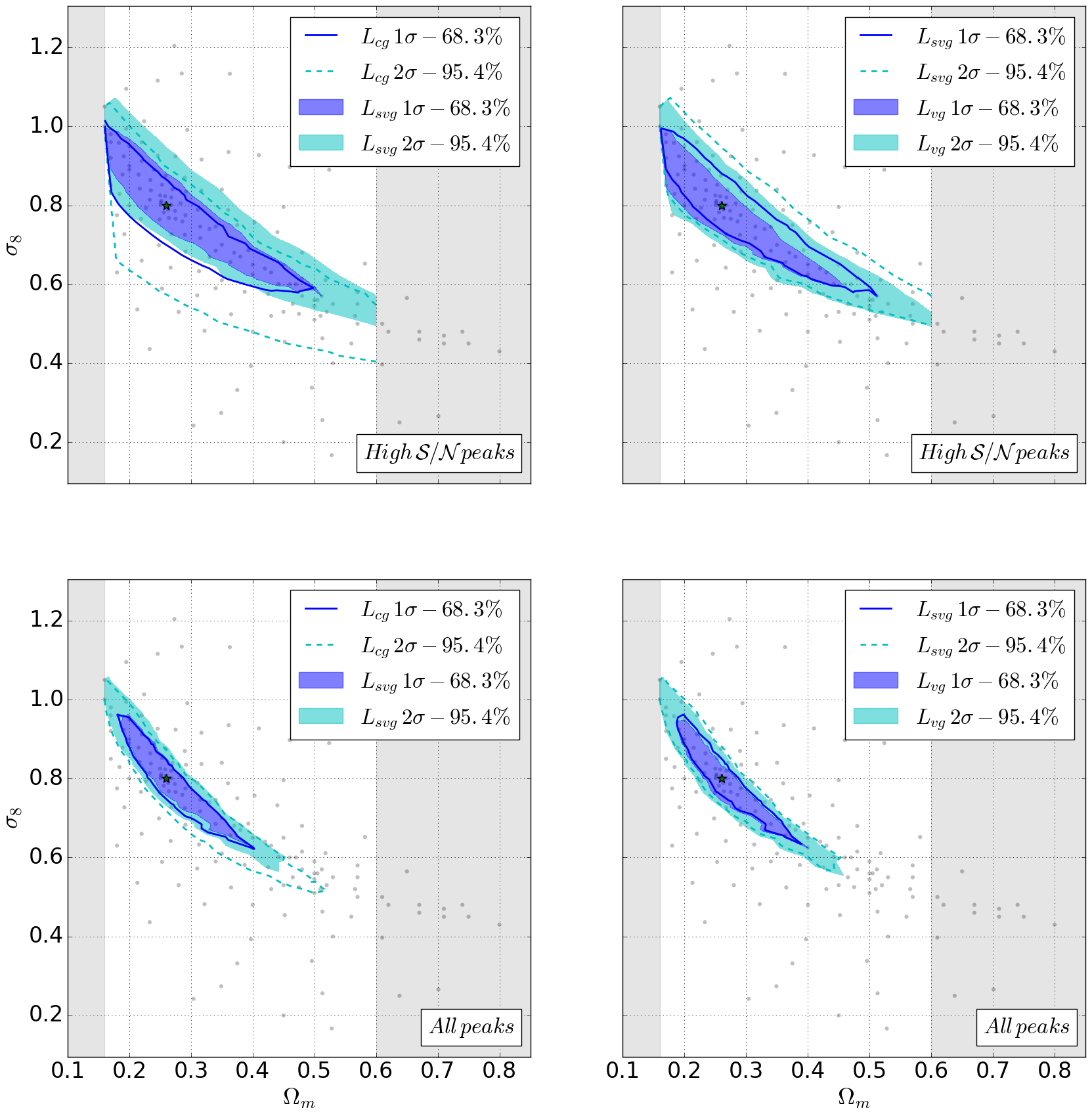}
\end{center}
\caption{Effect on the credibility regions of using a cosmology-dependent covariance. In the left panels we show the change caused by introducing a variable covariance in the $\chi^2$ term of a Gaussian likelihood ($L_{svg}$, shaded areas) compared with a constant covariance ($L_{cg}$, lines).  On the right we display the change from using a variable covariance matrix in the determinant term as well ($L_{vg}$, shaded areas) compared with $L_{svg}$ (lines). The upper panels show the result using only high-significance $(\mathcal{S/N}>3)$ peaks, while the lower panels show results with all peaks included. Introducing a variable covariance in the $\chi^2$ has a larger impact than using it in the determinant term. Also, the effects are larger when using only high significance peaks (see Table~\ref{cosmodep}).}
\label{contours2}
\end{figure*}

Finally, we assessed the impact of using a variable covariance matrix when computing the likelihood in the same way as was done in \cite{LKII}. Estimating the covariance at each point of the parameter space is computationally expensive, but as we have shown, the covariance can change significantly. Fig.~\ref{contours2} shows the effect on both $1\sigma$ (68.3\%) and $2\sigma$ (95.4\%) contours; the values for the changes are listed in Table \ref{cosmodep}. The effects are always more important if only high-significance peaks are included. Introducing a variable covariance in the $\chi^2$ term of a Gaussian likelihood --i.e., using $L_{svg}$ instead of $L_{cg}$-- tightens constraints by $14-19\%$ ($14-32\%$ for high $\mathcal{S/N}$ peaks only). Incorporating it also to the determinant term --i.e., going from $L_{svg}$ to $L_{vg}$-- has a more limited impact of $0-1\%$ ($13-19\%$ for high $\mathcal{S/N}$ peaks-only). It would be advisable then to use a cosmology-dependent covariance for a precise determination of parameter constraints, with the exception of those cases in which most of the parameter space has been rejected by previous experiments and only a small region needs to be explored.

\begin{table}
\caption{\label{cosmodep}Effect of using a cosmology-dependent covariance matrix. $1\sigma$ (68.3\%) and $2\sigma$ (95.4\%) credible contours are computed using the three likelihoods described in \ref{inference} ($L_{cg}$, $L_{svg}$ and $L_{vg}$). The analysis is done twice, using only high significance peaks $(\mathcal{S/N}>3)$ and all the peaks. We report the figure of merit (FoM); defined as the inverse of the area of the credibility region), changes in the credibility regions and shifts in their centroid.
Introducing a cosmology-dependent covariance into the $\chi^2$ term of the Gaussian likelihood has a bigger impact than introducing it in the determinant term. Also, the effect is bigger when only high peaks are included.}
\begin{ruledtabular}
\begin{tabular}{lcccc}
$\mathcal{S/N}>3$ Peaks\\
\hline
\hline
Likelihood & FoM & $\Delta\,Area$ & $\Delta \Omega_m$ & $\Delta \sigma_8$ \\
\hline
$1\sigma\,L_{cg}$ 	& 25 		& - 		& - 		& - 		\\
$1\sigma\,L_{svg}$ 	& 29 		& -14\% 	& +0.01 	& +0.02 	\\
$1\sigma\,L_{vg}$ 	& 36	 	& -19\% 	& -0.03 	& +0.01 	\\
\hline
$2\sigma\,L_{cg}$ 	& 9 		& - 		& - 		& - 		\\
$2\sigma\,L_{svg}$ 	& 13 		& -32\% 	& -0.01 	& +0.09 	\\
$2\sigma\,L_{vg}$ 	& 15 		& -13\% 	& -0.01 	& -0.01 	\\
\hline
\hline
All peaks\\
\hline
\hline
Likelihood & FoM & $\Delta\,Area$ & $\Delta \Omega_m$ & $\Delta \sigma_8$ \\
\hline
$1\sigma\,L_{cg}$ 	& 69 		& - 		& - 		& - 		\\
$1\sigma\,L_{svg}$ 	& 81 		& -14\% 	& -0.00 	& +0.01 	\\
$1\sigma\,L_{vg}$ 	& 81	 	& -0\% 	& +0.00 	& -0.01 	\\
\hline
$2\sigma\,L_{cg}$ 	& 26 		& - 		& - 		& - 		\\
$2\sigma\,L_{svg}$ 	& 32 		& -19\% 	& -0.02 	& +0.05 	\\
$2\sigma\,L_{vg}$ 	& 32 		& +1\% 	& +0.00 	& -0.01 	\\
\end{tabular}
\end{ruledtabular}
\end{table}

%%%Figure

% Discussion****************************************************************************************************
\section{Discussion}\label{discussion}
Given the restricted scope of this paper --to assess the accuracy of the halo-based model \textsc{Camelus} for cosmological inference using WL peaks-- our main findings are the differences between its credible contours and those from N-body simulations. 

\begin{figure*}
\begin{center}
\includegraphics[width=1.0\textwidth]{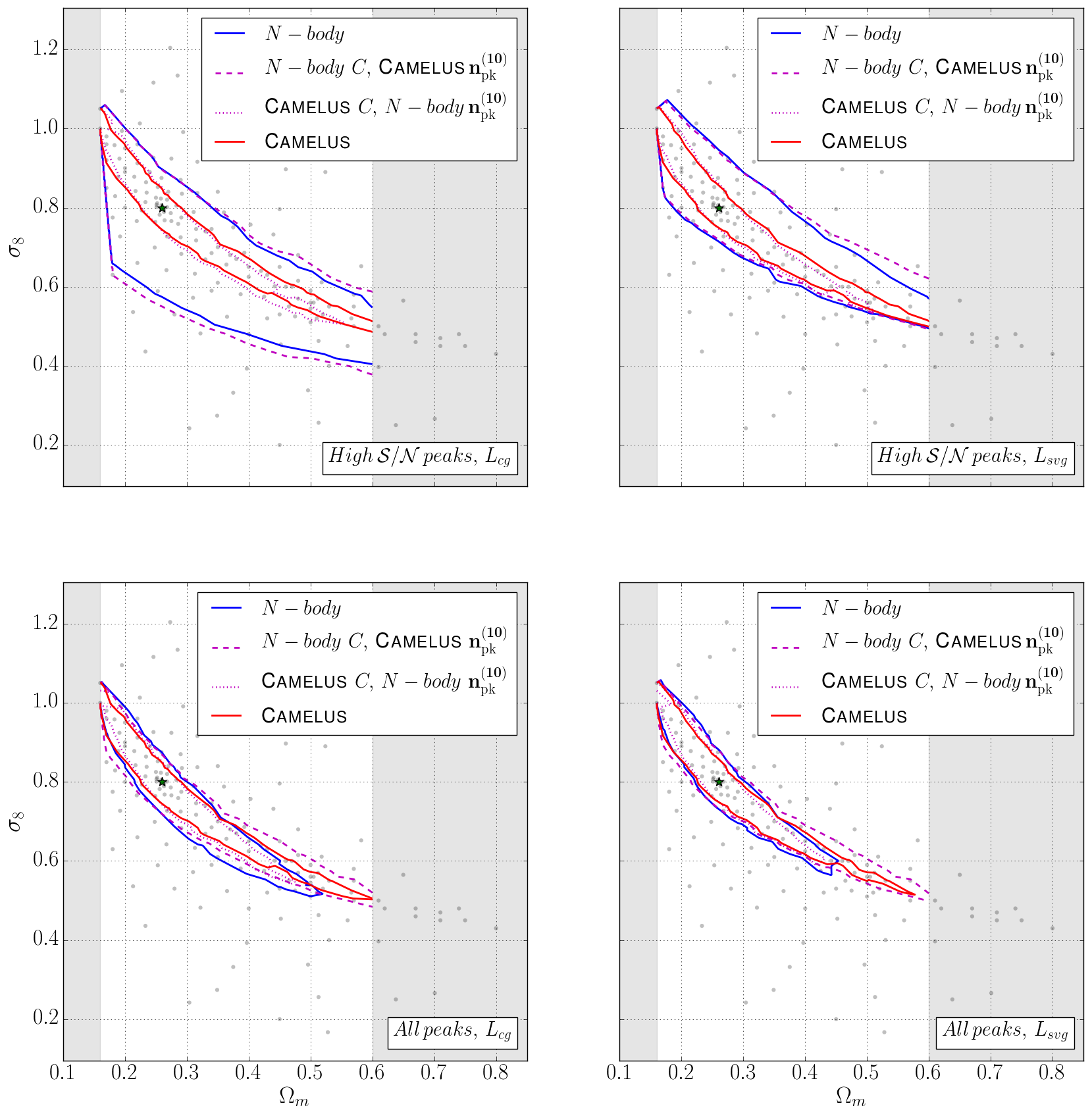}
\end{center}
\caption{Impact of differences in peak counts and covariance matrices on credible contours. Solid lines are $2\sigma$ contours from N-body (blue) and \textsc{Camelus} (red) data. Magenta lines are contours computed mixing peak counts from one model with the covariance matrices from the other. The dashed contours combine N-body covariance matrices with \textsc{Camelus} peak counts, and the dotted contours combine conversely N-body peak counts with \textsc{Camelus} covariances. The upper panels show the results using only $\mathcal{S/N}>3$ peaks while the lower panels display the contours obtained including all peaks. On the left we show contours computed using a constant covariance, $L_{cg}$ and on the right those introducing a variable covariance in the $\chi^2$ term, $L_{svg}$. 
In general, contours computed with the same covariance matrices are closer than those calculated with the same peak counts. The effect is more noticeable for the cases which include only high-significance peaks, since for these the N-body and \textsc{Camelus} contours exhibit a greater difference.}
\label{mix}
\end{figure*}

We identified small discrepancies in peak counts and significantly larger covariances from N-body data, with a stronger dependence on cosmology. To disentangle the effect of both elements on parameter inference, we computed "hybrid" likelihoods mixing peak counts from one model with covariance matrices from the other. Fig.~\ref{mix} shows the resulting $2\sigma$ credibility regions. Substituting the covariance for that from \textsc{Camelus} data shrinks the N-body contours to a thickness equivalent to that of \textsc{Camelus}. The effect on the credibility region from \textsc{Camelus} of using peak counts from N-body simulations is comparatively less important, suggesting that more accurate estimation of  covariances have the highest potential for improvement. The upper panels of Fig.~\ref{mix} were plotted using only high significance peaks and show even more clearly how differences in the covariance matrices drive the size and shape of the credible contours.

%%%Figure

\begin{figure}
\begin{center}
\includegraphics[width=0.5\textwidth]{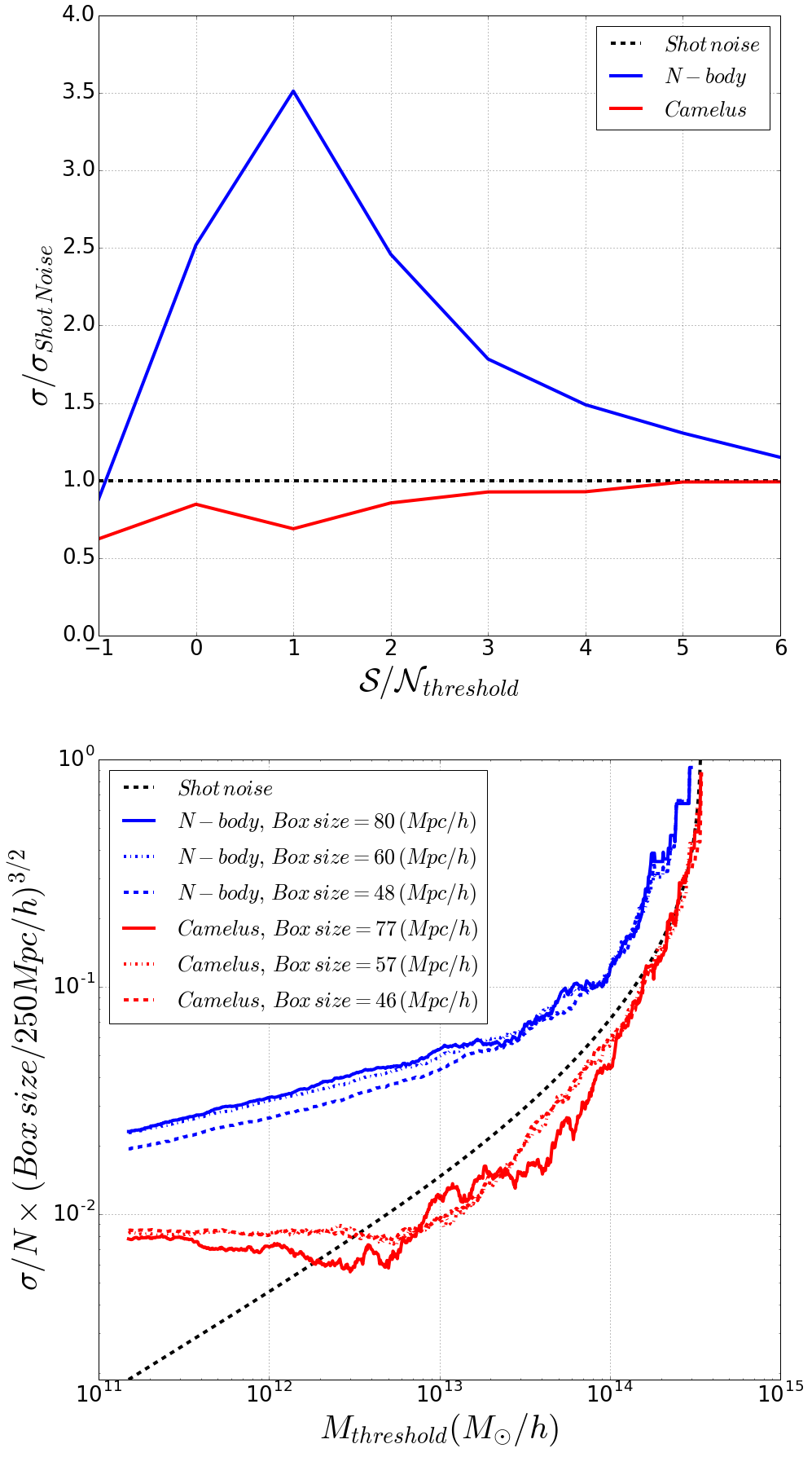}
\end{center}
\caption{Peak and halo count variance comparison between N-body (blue) and \textsc{Camelus} (red). 
\textbf{Upper panel:} ratio of the cumulative peak count standard deviation from its value  expected for a Poisson distribution, as a function of peak height. For pure Poisson shot noise, this ratio is unity (horizontal black dashed line). We find significantly higher sample variance than the results from \textsc{Camelus}, and what would be expected for a Poisson distribution. As the peak height increases and the peak counts decrease, shot noise starts to dominate.
\textbf{Lower panel:} variance of the cumulative halo number as a function of minimum halo mass. Sample variance is estimated from different sub-volumes, and scaled to a common reference volume of   $(250\, h^{-1} {\rm Mpc})^3$. Shot noise is estimated from a Poisson distribution with mean value adopted from a theoretical halo mass function \cite{Jenkins2001}. N-body cumulative halo counts exhibit a sample variance higher than expected from a Poisson distribution. Shot noise becomes more important at higher masses, as the halos become scarcer. \textsc{Camelus} is dominated by shot noise.}
\label{SampleVariance}
\end{figure}

To understand the origin of the discrepancy in peak-count variance, we compared halo counts from both models, since there is an established connection between halos and convergence peaks (\cite{Liu2016}). To identify halos in our N-body simulation we used the Amiga Halo Finder (AHF) \cite{AHF}. Since we evolved a single $240\,h^{-1} {\rm Mpc}$ box per cosmology, we subdivided it into sub-volumes to compute the variance. We split our simulation volume in $3^3$, $4^3$ and $5^3$ equally sized sub-boxes and scaled the counts to a common reference volume. We ran \textsc{Camelus} to generate halo catalogues corresponding to similar volumes as those of the sub-boxes used for the N-body calculation, and scaled the counts in the same way. The results are shown in Fig.~\ref{SampleVariance} and are in good agreement with analogous findings for cluster counts \cite{Hu2003}. Cumulative halo counts from N-body simulations have a higher sample (cosmic) variance than what would be expected if it were due solely to shot noise that follows a Poisson distribution. We use \cite{Jenkins2001} for the mean counts in the shot noise calculation. This is the same halo mass function used in \textsc{Camelus}, and we verified that it was in good agreement with the halos extracted from our N-body simulation. The excess sample variance is caused by LSS clustering halos which increases the correlation of their positions. As halos become more massive and rarer, shot noise becomes more important and the excess sample variance diminishes. 

\textsc{Camelus} places halos randomly, and its halo sample variance is dominated by shot noise except for the low-mass tail of the halo distribution. Halos are sampled from an analytical mass function until the total mass in a volume reaches its expected mean value. This condition that the total mass in halos is fixed links high-- and low-mass halo numbers, transferring variance to the low-mass halo range. Nevertheless, this effect does not translate into larger covariances, since low-mass halos do not contribute to peak counts. We compared peak counts from \textsc{Camelus} using different minimum halo masses ($10^{10}$, $10^{11}$  and $10^{12}\,\rm{M_{\odot}}$) and found virtually no difference.

Convergence peaks resulting from the projected mass density field, exhibit a similar pattern. The upper panel of Fig.~\ref{SampleVariance} shows the variance in the cumulative peak counts as a function of their height. Peak counts from N-body data also have a higher sample variance compared to a Poisson distribution and, as the peak $\mathcal{S/N}$ increases, shot noise becomes more important. For \textsc{Camelus} data, sample variance is smaller and is dominated by shot noise. The counts come from 500 $3.5\times3.5\,{\rm deg^2}$ convergence maps for the fiducial cosmology.

The parallel between halo and peak-count sample variance suggests that modifying the \textsc{Camelus} algorithm to account for halo clustering could enhance its accuracy by yielding larger covariance matrices that would propagate into looser parameter constraints.

%%%Figure

\begin{figure}
\begin{center}
\includegraphics[width=0.5\textwidth]{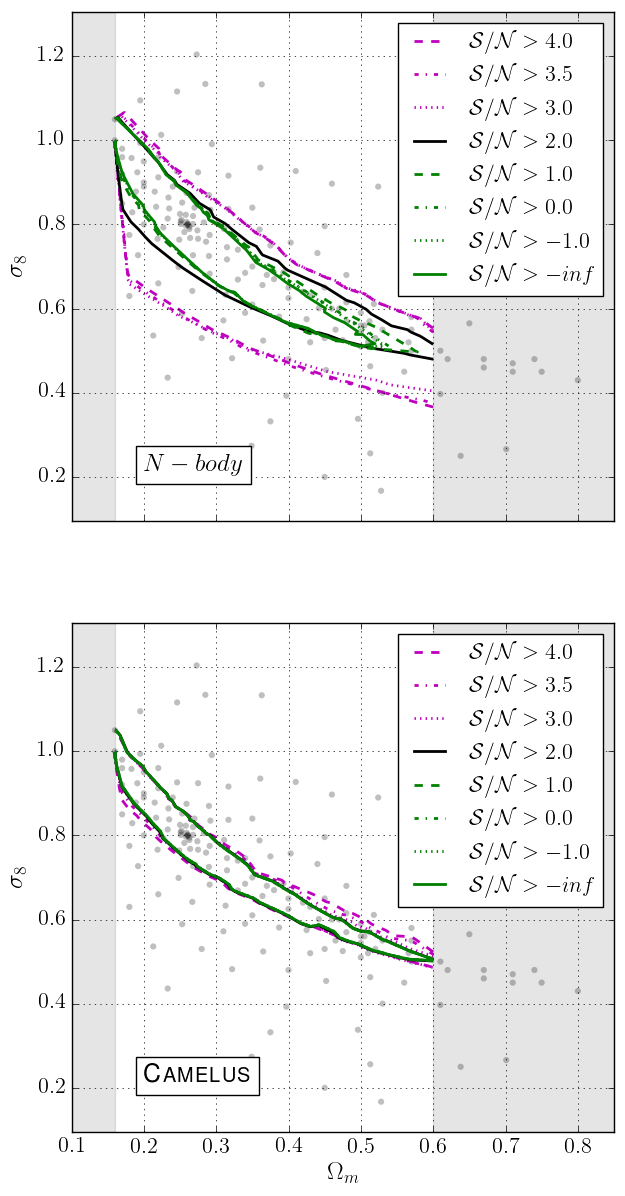}
\end{center}
\caption{Influence on the credibility region of the lowest significant peaks included in the $(L_{cg})$ likelihood calculation. \textbf{Upper panel:} for N-body simulations, including peaks with $2.0<\mathcal{S/N}<3.0$ significantly improves the model's predictive power. \textbf{Lower panel:} for \textsc{Camelus}, little or no improvement in predictive power is found when lower-significance peaks are included.}
\label{bineffect}
\end{figure}

We also found that including low-significance peaks in the analysis improves the predictive power for N-body simulations, while it does little for \textsc{Camelus}. Fig.~\ref{bineffect} and Table~\ref{bineffectTable} show the effect of adding bins of decreasing significance peaks to the contours' computation. For N-body simulations, the impact is particularly important when peaks in the range $\mathcal{S/N}\in[2,3]$ are incorporated, with $2\sigma$ contours reduced by $25-48\%$. Those moderately low-significance peaks have been associated with constellations of small halos (\cite{Yang2011, Liu2016}). These alignments are missing in the halo catalogs generated with \textsc{Camelus}, which constrains cosmology essentially through high peaks which are caused by high-mass halos.

%%%Figure

\begin{table}
\caption{\label{bineffectTable}Impact on the models' predictive power of the lowest significance peak bin included in analysis. Figure of merit (FoM) and change in $2\sigma$ contour area ($\Delta\%$) for constant, semi-varying and variable covariance likelihoods.}
\begin{ruledtabular}
\begin{tabular}{l|cc|cc|cc}
$N-body$ & \multicolumn{2}{c}{$L_{cg}$} & \multicolumn{2}{c}{$L_{svg}$} & \multicolumn{2}{c}{$L_{vg}$} \\
\hline
& FoM & $\Delta(\%)$ & FoM & $\Delta(\%)$ & FoM & $\Delta(\%)$ \\
\hline
$\mathcal{S/N}>4.0$ & 9 		& - 		& 11 		& - 		& 13 		& - \\
$\mathcal{S/N}>3.5$ & 8 		& -0 		& 12 		& -2 		& 14 		& -4 \\
$\mathcal{S/N}>3.0$ & 9 		& -4 		& 13 		& -11 	& 15 		& -9 \\
$\mathcal{S/N}>2.0$ & 17 	& -48 	& 19 		& -32 	& 20 		& -25 \\
$\mathcal{S/N}>1.0$ & 22 	& -22 	& 27 		& -31 	& 27 		& -27 \\
$\mathcal{S/N}>0.0$ & 24 	& -10 	& 31 		& -11 	& 30 		& -10 \\
$\mathcal{S/N}>-1.0$ & 25 	& -3 		& 31 		& -1 		& 31 		& -1 \\
$\mathcal{S/N}>-\inf$ & 26 	& -6 		& 32 		& -3 		& 32 		& -4 \\
\hline \hline
\textsc{Camelus} & \multicolumn{2}{c}{$L_{cg}$} & \multicolumn{2}{c}{$L_{svg}$} & \multicolumn{2}{c}{$L_{vg}$} \\
\hline
& FoM & $\Delta(\%)$ & FoM & $\Delta(\%)$ & FoM & $\Delta(\%)$ \\
\hline
$\mathcal{S/N}>4.0$ & 26  	& -     	& 27 		& - 		& 27 		& - \\
$\mathcal{S/N}>3.5$ & 30  	& -12 	& 30 		& -10 	& 30 		& -10 \\
$\mathcal{S/N}>3.0$ & 33  	& -10	& 33 		& -9	 	& 34 		& -9\\
$\mathcal{S/N}>2.0$ & 35  	& -4	 	& 36 		& -8 		& 36 		& -8 \\
$\mathcal{S/N}>1.0$ & 36  	& -3	 	& 37 		& -2 		& 37 		& -2 \\
$\mathcal{S/N}>0.0$ & 36  	& -1	 	& 37 		& -1 		& 37 		& -1 \\
$\mathcal{S/N}>-1.0$ & 36 	& +0 	& 37 		& -0 		& 38 		& -0 \\
$\mathcal{S/N}>-\inf$ & 36 	& -0 		& 38 		& -0 		& 38 		& -1 \\
\end{tabular}
\end{ruledtabular}
\end{table}

Our likelihood calculations rely on a precise estimation of the precision matrix, $C^{-1}$, and the determination of the credible contours on the interpolation of the likelihood beyond the discrete set of cosmologies for which we run simulations.

%%%Figure
\begin{figure}
\begin{center}
\includegraphics[width=0.5\textwidth]{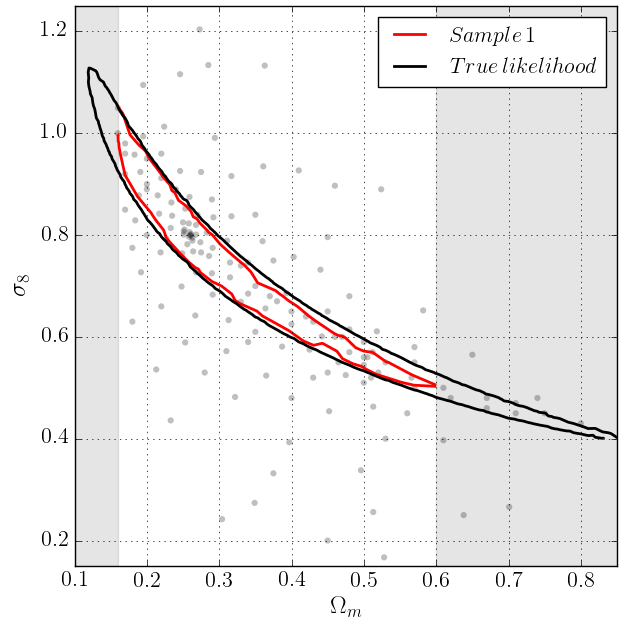}
\end{center}
\caption{Effect on \textsc{Camelus} credible contours of finite sampling of the cosmological parameter space. $2\sigma$ contours obtained from a fine grid of 7,803 models (black) and interpolated from our suite of 162 cosmologies (red). The interpolated contour is smaller in the low- and high-$\Omega_m$ tails. Thus we excluded from our analyses the greyed-out regions, corresponding to $\Omega_m<0.160$ and $\Omega_m>0.600$.}
\label{sampling}
\end{figure}

%%%Figure
\begin{figure}
\begin{center}
\includegraphics[width=0.5\textwidth]{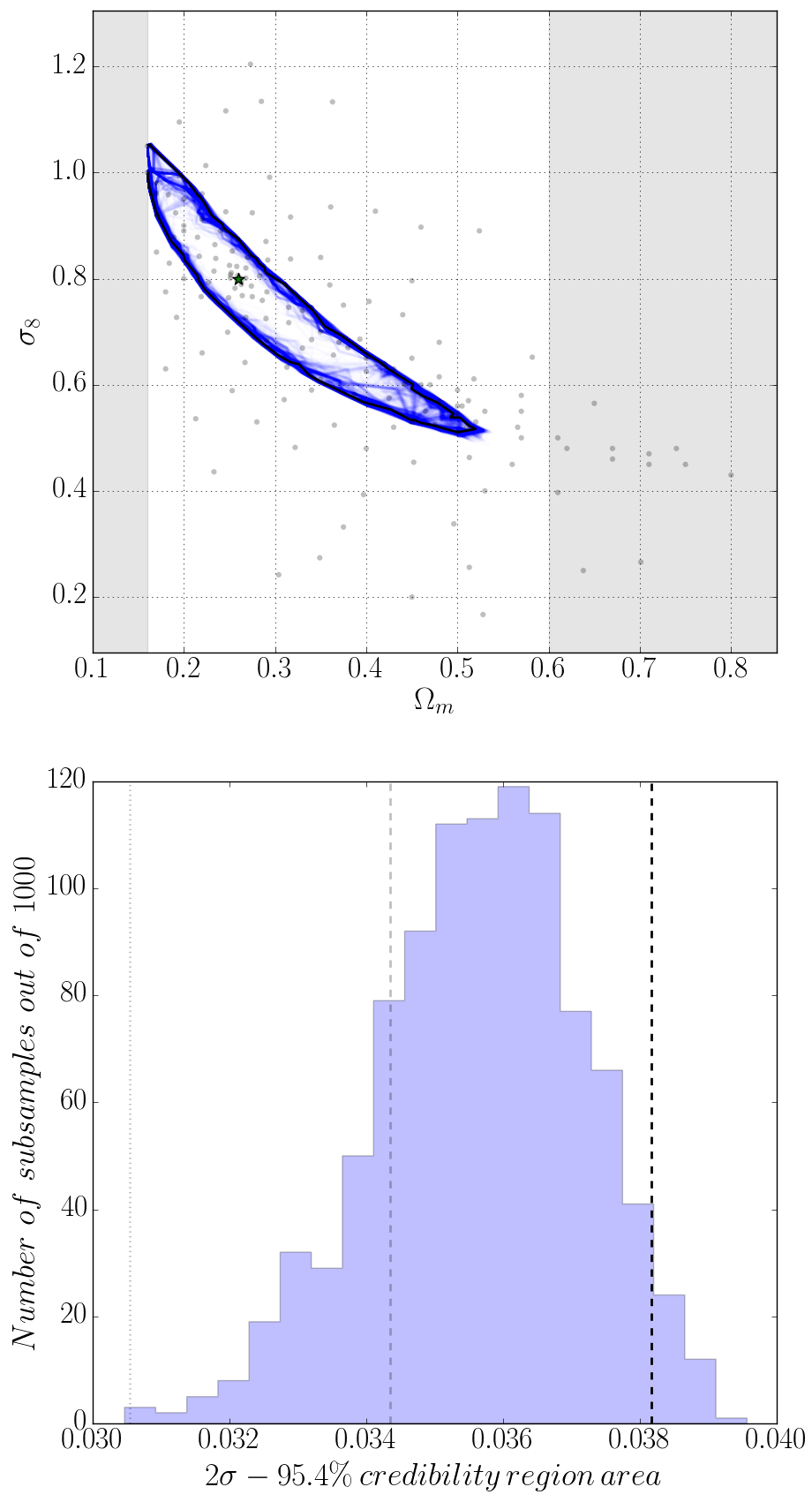}
\end{center}
\caption{Effect of cosmological parameter sampling on the N-body credible contours. We draw $1,000$ bootstrap samples from our suite of cosmologies; i.e., we draw samples of 162 elements with substitution, each having on average 102-103 different cosmologies. \textbf{Upper panel:} $L_{cg}$ $2\sigma$ contours from the full suite (black) and the 1,000 sub-samples (blue). Darker areas indicate higher contour concentration. \textbf{Lower panel:} area histogram for the bootstrap samples. Displayed for reference are the area for the full suite (black dashed line), 90\% of this value (grey dashed line) and 80\% (grey dotted line). 81\% of the contours fall within 10\% of the original area and 99\% within 20\%.}
\label{bootstrap}
\end{figure}

For each cosmology, we estimated the covariance matrices using 500 converge field realizations recycled from a single N-body calculation by slicing, shifting and rotating the simulated box. Previous work showed (\cite{PetriSampleVar}) that a single N-body run is sufficient to generate $\approx10^4$ convergence maps whose peak counts are statistically independent, and two boxes would be enough to measure feature means with an accuracy of $50\%$ of the statistical error. Therefore, we decided to use a single box, which allowed us to maximize the number of cosmologies to sample given our available computing resources. 

While the inverse of a covariance matrix estimated from data is not unbiased, since the number of realizations we use (500) is much larger than the dimension of our data vectors (10), the bias is negligible $(\approx 2 \%)$. We verified that the results with Gaussian likelihoods after de-biasing the covariances following \cite{Hartlap2007} were the same as those from using the non-Gaussian form of the likelihood found in \cite{Sellentin}.

Interpolation can also introduce errors in the contours. We verified this effect on the \textsc{Camelus} contours by running an additional fine grid of 7,803 cosmologies --described in Fig. 1 of ref.~\cite{LKII}--, and plotting the contours obtained from these and our original models in Fig.~\ref{sampling}. The regions corresponding to low- and high-$\Omega_m$ values are under-sampled, and as a result the contours in those regions are underestimated. Therefore, we limited our contour analyses to the interval $\Omega_m \in [0.160,0.600]$, where the true and the estimated contours agree within 20\%. 

Since we could not reproduce this analysis for our N-body simulations due to the computational cost, we generated contours from bootstrap samples of our full simulation set. That is, we 
sampled from the 162 cosmologies, with substitution, and drew the resulting contours in Fig. ~\ref{bootstrap}. Each sample had an average of 102-103 unique cosmologies in them. As with the analysis of the effect of sampling on the \textsc{Camelus} contour, we found that dropping models almost always results in a smaller area, and as a result we may be underestimating the errors on the parameters. We expect that underestimation to be moderate, for $81\%$ of the samples yield areas that lie within $10\%$ of the area computed with the full sample and $99\%$ of the samples fall within $20\%$. The highest risk is missing part of the tail of the credibility region, which occurs in some of the random bootstrap samplings. 

We do not address the question of whether a Gaussian likelihood is an appropriate model for our data, since the focus of this study is to compare the results from the two models. We will treat it in future work. For \textsc{Camelus} data, the Gaussian approximation yields credible contours in good agreement with those computed using the actual distribution of peak counts, as can be seen in the left panel of Fig. 8 in \cite{LKII}.

Other underlying simplifications common to both the N-body and halo-based simulations used in this work are the non-inclusion of baryonic effects, the Born and flat sky approximations, and the omission of any survey effects such as masking, instrument systematics, etc. Baryons have been shown to increase the amplitude of the WL power spectrum on small scales and to introduce a small bias in high $\mathcal{S/N}$ peaks \cite{Yang2013}. The precision requirements and large sky coverage from future surveys will require the inclusion of these baryonic effects \cite{Weinberg}, as well as revisiting some of the approximations used in our models \cite{Kilbinger15}.

In future work, new modified ray-tracing simulations using manipulated snapshots from N-body simulations may clarify the specific sources of discrepancy between N-body and halo-based models. Possible reasons can be enumerated as follows:

\begin{enumerate}[(i)]
\item Non-halo contributions, e.g. filaments, walls,
\item Halo clustering,
\item Non NFW halo profiles, e.g. merging halos, triaxiality, and
\item Halo concentration; e.g., broad distribution instead of a deterministic function.
\end{enumerate}

Modifications to a model such as \textsc{Camelus} to address points (ii)-(iv) could in principle be addressed within the halo model framework and would make it even more useful as a fast lensing emulator by improving its accuracy.

% Conclusions**************************************************************************************************
\pagebreak
\section{Conclusions}\label{conclusions}
In this work we compared the outcomes from the fast halo-based algorithm \textsc{Camelus} with those of N-body simulations for a suite of cosmologies spanning a wide range of values in the $\{\Omega_m, \sigma_8\}$ plane.

We found larger (by $\approx30\%$ in area), more significantly tilted (by $\approx 13\%$ in angle) credible contours from N-body data. Importantly, the two models draw their predictive power from a different types of peaks. While \textsc{Camelus} constrains cosmology through high--$\mathcal{S/N}$ peaks associated with massive halos, the N-body data are highly sensitive to lower-$\mathcal{S/N}$ peaks.

The larger thickness and overall area of the N-body credible contours are mostly driven by the covariances, with peak counts showing a higher variance than expected from pure shot noise. This suggest that modifying the placement of halos in \textsc{Camelus} to account for the correlations in their locations is a promising way to improve its covariance estimation and accuracy as a WL peak count emulator.

Using a cosmology-dependent covariance matrix for likelihood estimation improves constraints by  $14-20\%$, and thus will be needed in order to achieve high-precision parameter estimations.

Finally, we have found that optimal sampling of a high-dimensional parameter space with expensive N-body simulations to define credibility regions with high precision is a topic that requires further investigation, and a fast simulator like \textsc{Camelus} could prove itself particularly valuable by providing a first estimation of the likelihood.

% Acknowledgements***************************************************************************************
\section*{Acknowledgments}
We thank Chieh-An Lin and Martin Kilbinger for useful discussions, comments on the manuscript, and their help with the use of their \textsc{Camelus} code. The simulations were performed at the NSF XSEDE facility, and at the Yeti computing cluster at Columbia University. This work was supported in part by the NSF Grant No. GG-008775 (to Z.H.) and by the Research Opportunities and Approaches to Data Science (ROADS) program at the Institute for Data Sciences and Engineering at Columbia University (to Z.H. and D.H.).  Z.H. also gratefully acknowledges support from a Simons Fellowship in Theoretical Physics.

\bibliography{paper.bib}

\end{document}